\documentclass[sigconf]{acmart}

\usepackage{booktabs} 
\usepackage{arydshln}

\acmDOI{10.475/123_4}

\acmConference[WWW'19]{ACM Web conference}{May 13-17, 2019}{San Francisco, California, The USA}
\acmYear{2019}
\copyrightyear{2019}

\usepackage{bm}
\usepackage{tikz}
\usepackage{algorithm}
\usepackage{algorithmic}
\usepackage{graphicx}
\usepackage{subfigure}
\usepackage{caption}
\usepackage{epstopdf}
\usepackage{multirow}
\usepackage{mathtools}
\def \VDMF {NVHCF}

\setcopyright{none}

\begin{document}
\title{Neural Variational Hybrid Collaborative Filtering}

\author{Teng Xiao}
\affiliation{
  \institution{Sun Yat-Sen University}
}
\email{xiaot25@mail2.sysu.edu.cn}

\author{Shangsong Liang}
\affiliation{
  \institution{Sun Yat-Sen University}
}
\email{liangshs5@mail.sysu.edu.cn}

\author{Hong Shen}
\affiliation{
  \institution{Sun Yat-Sen University}
}
\email{shenh3@mail.sysu.edu.cn}
\author{Zaiqiao Meng}
\affiliation{
  \institution{Sun Yat-Sen University}
}
\email{zqmeng@aliyun.com}

\begin{abstract}
Collaborative Filtering (CF) is one of the most widely used methods for Recommender System. Because of the Bayesian nature and non-linearity, deep generative models, e.g. Variational Autoencoder (VAE),  have been applied into CF task, and have achieved great performance. However, most VAE-based methods suffer from matrix sparsity and consider the prior of users'  latent factors to be the same, which leads to poor latent representations of users and items. 
Additionally, most existing methods model latent factors of users only and but not items, which makes them not be able to recommend items to a new user. 
To tackle these problems, we propose a Neural Variational Hybrid Collaborative Filtering, \VDMF{}. Specifically, we consider both the generative processes of users and items, and the prior of latent factors of users and items to be \emph{side ~information-specific}, which enables our model to alleviate matrix sparsity and learn better latent representations of users and items. For inference purpose, we derived a Stochastic Gradient Variational Bayes (SGVB) algorithm  to analytically approximate the intractable distributions of latent factors of users and items. Experiments conducted on two large datasets have shown  our method  significantly outperforms the state-of-the-art CF methods, including the VAE-based methods.
\end{abstract}

%
%

\maketitle

\section{Introduction}
Recommendation system (RS) is of paramount importance in social networks and e-commerce platforms. For instance, about 60\% of videos recommended in YouTube receive clicks~\cite{Davidson:2010:YVR:1864708.1864770}. RS aims at inferring users' preferences over items by utilizing their previous interactions. Collaborative Filtering (CF) is one of the most used approaches. Most traditional CF methods are based on  matrix factorization (MF)~\cite{mnih2008probabilistic,salakhutdinov2008bayesian}. However, these methods suffer from matrix sparsity problem and can not capture the non-linearity relationships between users and items. To tackle matrix sparsity problem, many  CF methods such as hybrid CF methods that incorporate side information, i.e., users' features  and items' content information into traditional MF. To extract more latent factors of side information,  some previous work utilizes different models, e.g., Latent Dirichlet Allocation (LDA) \cite{Wang2011Collaborative}, denoising auto-encoder~\cite{Wang2015Collaborative} and marginalized denoising auto-encoder \cite{Li2015Deep} to model side information of users and items. However, as discussed in \cite{He:2017:NCF:3038912.3052569}, these methods use inner product to model  interactions between users and items, which limits them to be powerful to capture non-linearity. To model non-linear interaction, many methods directly  use neural networks  to model  these interactions, such as Neural Collaborative Filtering (NCF)~\cite{He:2017:NCF:3038912.3052569}, Neural Factorization Machine (NFM)~\cite{He2017Neural} and DeepFM~\cite{DBLP:conf/ijcai/GuoTYLH17}, which have shown  promising  performance. However, these neural network-based models are deterministic, and can not capture the uncertainty of the users' and items' latent representations. \par
Because of the power of capturing uncertainty and the non-linearity of deep generative models \cite{Kingma2014Auto}, some recent methods such as VAE-CF \cite{Liang:2018:VAC:3178876.3186150}, Collaborative Variational Autoencoder (CVAE)~\cite{Li2017Collaborative} and have applied deep generative models such as Variational Autoencoder (VAE) \cite{Kingma2014Auto} into  CF task. Despite the effectiveness of these methods for CF, they demonstrate a number of drawbacks:  (1) For \cite{Liang:2018:VAC:3178876.3186150}, it only utilizes user-item feedback matrix, resulting in poor performance when the matrix is very sparse. (2) They worked through modeling users' behavior, which makes them can not  recommend an item to a new user. (3) They choose the same Gaussian prior for all users, 
which leads to very  poor latent representations of users and items \cite{hoffman2016elbo}.
(4) For \cite{Li2017Collaborative}, it directly uses inner product to model interaction hinders itself to learn non-linear interactions between users and items. \par
Accordingly, we solve the aforementioned problems by proposing a unified Neural Variational Hybrid Collaborative Filtering (\VDMF {}) for hybrid CF. 
Unlike many existing VAE-based methods that model users' or items' generative process, we model the generative process from the views of users and items through a unified conditional neural variational model,  which enables it to still work well for a new user or a new item. We consider the priors of  latent factors of users and items to be conditioned on their side information through a neural network. The parameters of prior neural network are learned  from data, leading to the fact that it is able to embed users' better preferences and items' features into latent factors of users and items, respectively, and alleviate matrix sparsity problem. For inferring the posterior of latent factors of users and items, we derived a Stochastic Gradient Variational Bayes (SGVB) algorithm to infer these posterior, which makes the parameters of our model can be effectively learned by back-propagation.
To sum up, our main contributions are as follows: \\
(1) We proposed a novel conditional neural variational framework to effectively learn nonlinear latent representations of users and items for hybrid CF. To the best of our knowledge, we are the first to model both users' and items' generative process through a unified conditional deep generative model. \\
(2) We incorporated side information of users and items into their latent factors through a conditional prior ways, which makes our model can alleviate matrix sparsity and cold start problems and model better latent representations of users and items.\\
(3) We derived tractable variational evidence lower bounds for our proposed model and devised a neural network to infer latent factors of users and items.\\
(4) We systematically conducted experiments on three public  datasets. Experimental results showed that our \VDMF {} model outperforms state-of-the-art CF methods.

\section{Related Work}
In recent years, deep learning has achieved tremendous achievements in various fields~\cite{lecun2015deep,krizhevsky2012imagenet}. Due to the the powerful abilities of neural networks to discover non-linear, subtle relationships in complex data for CF, many researchers utilize neural networks to address the task of CF. To incorporate item content information into latent factors of items, \citeauthor{Wang2015Collaborative} proposed collaborative deep learning (CDL)~\cite{Wang2015Collaborative} to integrate stacked denoising autoencoder (SDAE) into probabilistic matrix factorization (MF). \citeauthor{Li2015Deep} proposed Deep Collaborative Filtering Framework \cite{Li2015Deep}, which is a general framework for unifying deep learning approaches with CF. Recently, \citeauthor{dong2017hybrid} proposed the additional stacked denoising autoencoder (aSDAE)~\cite{dong2017hybrid} to incorporate side information into MF. \citeauthor{ijcai2017-447} proposed a novel matrix factorization model (DMF)~\cite{ijcai2017-447} with a neural network architecture. 
Since the above methods use inner product to model the interaction of  users and items, they are not able to capture the complex structure of the interaction data between users and items.  \citeauthor{He:2017:NCF:3038912.3052569} proposed Neural Collaborative Filtering (NCF)~\cite{He:2017:NCF:3038912.3052569}, which uses neural network to model interaction between users and items. \citeauthor{He2017Neural} proposed Neural Factorization Machines~\cite{He2017Neural}, which use Bi-Interaction layer to incorporate both feedback information and content information. See~\cite{zhang2017deep} for a more thorough
review of deep learning based recommender system.
Due to the power of capturing uncertainty and non-linearity of deep generative model~\cite{Kingma2014Auto}, some existing work utilizes deep generative models to address the task of CF.  \citeauthor{Li2017Collaborative} proposed Collaborative Variational Autoencoder (CVAE)~\cite{Li2017Collaborative}, which uses VAE to incorporate item content information into MF. \citeauthor{Liang:2018:VAC:3178876.3186150} directly utilize VAE structure~\cite{Liang:2018:VAC:3178876.3186150} for CF, and  \citeauthor{lee2017}  proposed a augmented VAE~\cite{lee2017} to incorporate auxiliary information to improve performance. Unlike previous VAE-based recommendation methods, we model the generative process of users and items through a unified neural variational framework, which makes our model to be able to  capture both users' and items' non-linear latent representations.


\section{Notations and Problem Definition}
We denote  user-item  feedback matrix by $\bm{R} \in \mathbb{R}^{M\times N}$, where $M$ and $N$ are the total number of users and items, respectively.
For implicit feedback, $R_{ij}$ = 1 indicates that
the $i$-th user has interacted with the $j$-th item and otherwise, $R_{ij}=0$. $\bm{F}\in \mathbb{R}^{P\times M}$ and $\bm{G}\in \mathbb{R}^{Q\times N}$ are the side information matrices of all users and items, respectively, with $P$ and $Q$ being the dimensions of each user's and item's side information, respectively. $ \bm{U}=[\bm{u}_{1},\ldots, \bm{u}_{M}]\in \mathbb{R}^{D\times M}$ and $\bm{V}=[\bm{v}_{1},\ldots, \bm{v}_{N}]\in \mathbb{R}^{D\times N}$ are the two rank matrices serving for users and items, respectively, with $D$ denoting the dimensions of latent factors. For convenient discussion, we represent each user $i$'s rating scores  over all items as $\bm{R}_{i\cdot}=[R_{i1},...,R_{iN}]\in \mathbb{R}^{N\times 1}$, where $R_{ij}$ is an element in $\bm{R}$. Similarly, we represent each item $j$'s rating scores from all users as $\bm{R}_{\cdot j}=[R_{1j},...,R_{Mj}]\in \mathbb{R}^{M\times 1}$. We call $\bm{R}_{i\cdot}$  and $\bm{R}_{\cdot j}$ as the collaborative information of user $i$ and item $j$, respectively.
Obviously, our task is to infer each user's and item's latent factors, $\bm{u}_{i}$ and $\bm{v}_{j}$ through $\bm{R}$, $\bm{F}$ and $\bm{G}$ to predict the missing $R_{ij}$.
\begin{figure}[H]
\centering
  \subfigure{
    \includegraphics[width=0.35\textwidth]{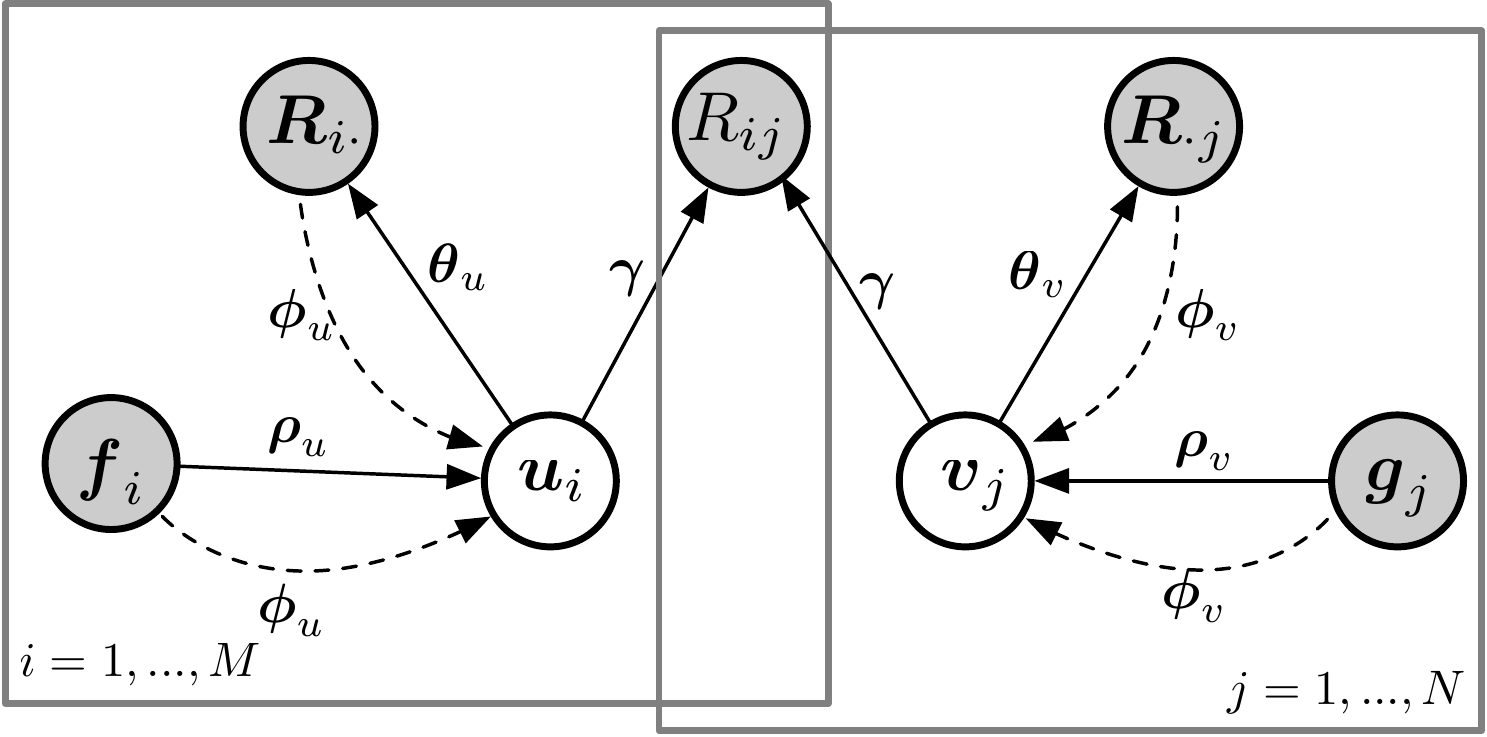}}
\caption{Graphical model of proposed \VDMF {}. Solid and dashed lines represent generative and inference process, respectively. Gray and white nodes represent  observed variables and latent variables, respectively. The symbols corresponding to lines denote the parameters of neural network.}
\label{subfig:PGM}
\end{figure}

\section{The Proposed Method}
\label{sec:vdmf}
In this section, we first present our neural variational hybrid collaborative filtering model, abbreviated as \VDMF{}, the probabilistic graphic model of which is shown in Figure ~\ref{subfig:PGM}.
\subsection{Neural Variational Hybrid Collaborative Filtering}
Most CF methods \cite{Liang:2018:VAC:3178876.3186150,wang2017irgan} assume that the prior distributions of user latent factor $\bm{u}_i$ and item latent factor $\bm{v}_j$ are standard  Gaussian distributions and predict rating only through user-item feedback matrix. Recently, incorporating different priors into VAE has achieved promising performance \cite{DBLP:conf/aaai/WangPVFZCRC18,Nalisnick_Smyth_2017}.
In our model, to further enhance the performance, besides the feedback matrix, we believe that the user's side information $\bm{f}_i$ can also positively contribute to the inference of his latent factor $\bm{u}_i$. Similarly, for better inferring the $j$-th item's latent factor $\bm{v}_j$, we also fully utilize the item's side information $\bm{g}_j$. Unlike most MF methods \cite{Wang2015Collaborative,Agarwal2009Regression,Kim2014Scalable} that incorporate side information via linear regression, in order to get more subtle latent relations and embed side information into latent factors of users and items, we consider that the conditional prior $p(\bm{u}_i|\bm{f}_i)$ and $p(\bm{v}_j|\bm{g}_j)$ are  \emph{side~information-specific} latent  Gaussian distributions such that we have $p(\bm{u}_i|\bm{f}_i)=\mathcal{N}(\bm{\mu} _u(\bm{f}_i), \bm{\Sigma}_u(\bm{f}_i))$ and $p(\bm{v}_j|\bm{g}_j)=\mathcal{N}(\bm{\mu}_v(\bm{g}_j), \bm{\Sigma}_v(\bm{g}_j))$, where
\begin{align}
&\bm{\mu}_u(\bm{f}_i)=F_{\bm{\mu}_u}(\bm{f}_i),\bm{\Sigma}_u(\bm{f}_i)=\mathrm{diag}(\exp(F_{\bm{\delta}_u}(\bm{f}_i))), \\
&\bm{\mu}_v(\bm{g}_j)=G_{\bm{\mu}_v}(\bm{g}_j),\bm{\Sigma}_v(\bm{g}_j)=\mathrm{diag}(\exp(G_{\bm{\delta}_v}(\bm{g}_j))).
\end{align}
Here $F_{\bm{\mu}_u}(\cdot)$, $F_{\bm{\delta}_u}(\cdot)$, are the two highly non-linear functions parameterized by $\bm{\mu}_u$ and $\bm{\delta}_u$ in the neural network, i.e., the user prior network, serving for all users, and  $G_{\bm{\mu}_v}(\cdot)$ and $G_{\bm{\delta}_v}(\cdot)$ are the two non-linear ones parameterized by  $\bm{\mu}_v$ and $\bm{\delta}_v$ in another neural network, i.e., the item prior network, serving for all items, respectively. For simplicity, note that we set $\bm{\rho}_u=\left \{ \bm{\mu}_u, \bm{\delta}_u \right \}$, $\bm{\rho}_v=\left \{ \bm{\mu}_v, \bm{\delta}_v \right \}$, $F_{\bm{\mu}_u}(\bm{f}_i)=\bm{\mu}_{\bm{u}_i}$, $\exp(F_{\bm{\delta}_u}(\bm{f}_i))=\bm{\delta}_{\bm{u}_i}$, $F_{\bm{\mu}_v}(\bm{g}_j)=\bm{\mu}_{\bm{v}_j}$ and $\exp(F_{\bm{\delta}_v}(\bm{g}_j))=\bm{\delta}_{\bm{v}_j}$.\par
For the $i$-th user's collaborative information, $\bm{R}_{i \cdot}$, we  believe that user's latent factor $\bm{u}_i$ can potentially affect user collaborative information $\bm{R}_{i \cdot}$. Then we consider  $\bm{R}_{i\cdot}$ to be generated from user latent factor $\bm{u}_i$, and governed by the parameter $\bm{\theta}_u$ in the generative network such that we
have:
\begin{equation}
\label{eq:aisiuN}
  \bm{R}_{i\cdot}\sim p_{\bm{\theta}_u}(\bm{R}_{i \cdot}|\bm{u}_i).
\end{equation}
Similarly, the $j$-th item's collaborative information, $\bm{R}_{\cdot j}$, is generated from item latent factor $\bm{v}_j$, and is governed by the parameter $\bm{\theta}_v$ in the generative network such that we have:
\begin{equation}
\label{eq:cjsivN}
\bm{R}_{\cdot j} \sim p_{\bm{\theta}_v}(\bm{R}_{\cdot j}|\bm{v}_j).
\end{equation}
Since our user-item matrix is implicit feedback matrix, the $\bm{R}_{i\cdot}$ and $\bm{R}_{\cdot j}$ are binary vectors. We consider the  $p_{\bm{\theta}_u}(\bm{R}_{i \cdot}|\bm{u}_i)$ and  $p_{\bm{\theta}_v}(\bm{R}_{\cdot j}|\bm{v}_j)$ to be  multivariate Bernoulli distribution.
For value $R_{ij}$, we consider it to be generated from $\bm{u}_i$ and $\bm{v}_j$ through a generative neural network parameterized by $\bm{\gamma}$:
\begin{equation}
R_{ij}\sim p_{\bm{\gamma}}(R_{ij}|\bm{u}_{i},\bm{v}_{j}).
\end{equation}
$R_{ij}$ is generated through a multi-layer perception network (MLP) parameterized by $\bm{\gamma}$.
Because of the one-class nature of implicit feedback \cite{Pan2008One}, we model $p_{\bm{\gamma}}(R_{ij}|\bm{u}_{i},\bm{v}_{j})$ as a Bernoulli distribution:
\begin{align}
\log p_{\bm{\gamma}}(R_{ij}|\bm{u}_{i},\bm{v}_{j})=
R_{ij}\log \hat{R}_{ij} +(1-R_{ij})\log(1-\hat{R}_{ij}), \label{Eq:cross-entropy}
\end{align}
where $\hat{R}_{ij}$ is the output of  the MLP.

According to the graphical representation of \VDMF{} at the right panel of Figure \ref{subfig:PGM}, the conditional joint distribution of \VDMF{} is factorized as: 
\begin{align}
p(\bm{R}, &~\bm{U},\bm{V}|\bm{F},\bm{G})=\prod_{i=1}^{M}\prod_{j=1}^{N}
\underbrace{p(\bm{u}_i|\bm{f_i})p(\bm{R}_{i\cdot}|\bm{u}_i)}_{\text{for users}}\cdot \displaybreak[3]\nonumber \\
&\underbrace{p(\bm{v}_j|\bm{g}_j)p(\bm{R}_{\cdot j}|\bm{v}_j)}_{\text{for items}}p(R_{ij}|\bm{u}_i,\bm{v}_j). \label{eq:jointDistribution}
\end{align}
\par
\noindent Instead of inferring the joint distribution, i.e., Eq.~\ref{eq:jointDistribution}, we are more interested in approximately inferring its posterior distributions over users' and items' factor matrices, $\bm{U}$, $\bm{V}$.  Traditional variational bayesian matrix factorization~\cite{Kim2014Scalable,park2013hierarchical} approximates the posterior distribution by using mean-field variational method, considers variation distribution that satisfies element-wise independence, and yields very good performance. However,  it is intractable to infer $\bm{U}$ and $\bm{V}$ by using traditional mean-field approximation since we do not have any conjugate probability distribution in our model which requires by traditional mean-field approaches. Inspired by VAE \cite{Kingma2014Auto}, we use Stochastic Gradient Variational Bayes (SGVB) estimator to approximate posteriors  of the latent variables related to user ($\bm{u}_i$)  and latent variables related to item ($\bm{v}_j$) by introducing two inference networks, i.e., the user inference network and the item inference network, parameterized by $\bm{\phi}_{u}$ and $\bm{\phi}_{v}$, respectively. To do this, we first decompose the variational distribution $q$ into two categories of variational distributions used in the two networks in our \VDMF{} model --- user inference network and item inference network, $q_{\bm{\phi}_{u}}$ and $q_{\bm{\phi}_{v}}$, by assuming the conditional independence:
\begin{align}
\label{VS}
& q(\bm{u}_i,\bm{v}_j|\bm{f}_i, \bm{R}_{i\cdot}, \bm{g}_j, \bm{R}_{\cdot j}, R_{ij})= q_{\bm{\phi}_{u}}(\bm{u}_i|\mathcal{X}_i) \cdot q_{\bm{\phi}_v}(\bm{v}_j|\mathcal{Y}_j),
\end{align}
where  $\mathcal{X}_i=\{ \bm{f}_i, \bm{R}_{i \cdot} \}$ and $\mathcal{Y}_j=\{\bm{g}_j, \bm{R}_{\cdot{j}}\}$.
Like VAE \cite{Kingma2014Auto}, the variation distributions are chosen to be a Gaussian distribution $\mathcal{N}(\bm{\mu},\bm{\Sigma})$, whose mean $\bm{\mu}$ and covariance matrix $\bm{\Sigma}$ are the output of the inference network. Thus, in our \VDMF{}, for latent variables related to the $i$-th user, we set:
\begin{align}
q_{\bm{\phi}_u}(\bm{u}_i|\mathcal{X}_i)=\mathcal{N}(\mu_{\bm{\phi}_{u}}(\mathcal{X}_i),\mathrm{diag}(\exp(\delta_{\bm{\phi}_{u}}(\mathcal{X}_i)))), \label{inference-u}
\end{align}
where the subscripts of $\mu$ and $\delta$ indicate the parameters in our user inference network corresponding to $\bm{u}_i$.  For simplicity, note that
we set $\mu_{\bm{\phi}_{u}}(\mathcal{X}_i)=\bm{\mu}_{\bm{u}_i}'$ and $\exp(\bm{\phi}_{u}(\mathcal{X}_i))=\bm{\delta}_{\bm{u}_i}'$, respectively.
Similarly, for latent variables related to the $j$-th item, we set:
\begin{align}
q_{\bm{\phi}_{v}}(\bm{v}_j|\mathcal{Y}_j)=\mathcal{N}(\mu_{\bm{\phi}_{v}}(\mathcal{Y}_j),\mathrm{diag}(\exp(\delta_{\bm{\phi}_{v}}(\mathcal{Y}_j)))), \label{inference-v}
\end{align}
where the subscripts of $\mu$ and $\delta$  indicate the parameters in item inference network corresponding to $\bm{v}_j$. For simplicity, note that
we set $\mu_{\bm{\phi}_{v}}(\mathcal{Y}_j)=\bm{\mu}_{\bm{v}_j}'$ and $\exp(\bm{\phi}_{v}(\mathcal{Y}_j))=\bm{\delta}_{\bm{v}_j}'$, respectively.

Thus, the tractable standard evidence lower bound (ELBO)~\cite{wainwright2008graphical}  for the inference can be computed as follows:
\begin{align}
\mathcal{L}&(q)=\mathbb{E}_q[\log p(\mathbf{R},\bm{U}, \bm{V}| \mathbf{F},\mathbf{G}) -\log q(\bm{U}, \bm{V}|\mathcal{O})] \displaybreak[3]  \label{Eq:fullloss} \\
&=\sum_{i=1}^{M}\sum_{j=1}^{N}(\mathcal{L}_{i}(q_{\bm{\phi}_u})+\mathcal{L}_{j}(q_{\bm{\phi}_v})+\mathbb{E}_q[\log p_{\bm{\gamma}}(R_{ij}|\bm{u}_i,\bm{v}_j)]),\nonumber
\end{align}
where $\mathcal{O}=(\bm{F},\bm{G},\bm{R})$ is a set of all observed variables. $q_{\bm{\phi}_{u}}$ and $q_{\bm{\phi}_{v}}$ are user term and item term in Eq.~\ref{VS}, respectively. Maximizing the ELBO is equivalent to use the variational distributions, i.e., $q_{{\bm{\phi}}_u}(\bm{u}_i|\mathcal{X}_i)$ and $q_{{\bm{\phi}}_v}(\bm{v}_j|\mathcal{Y}_i)$ to approximate their true posteriors ($p(\bm{u}_i|\mathcal{O})$ and $p(\bm{v}_j|\mathcal{O})$ ). For user $i$ and item $j$, we have:
\begin{align}
\mathcal{L}_{i}(q_{\bm{\phi}_u})=&\mathcal{L}(\bm{\phi}_u,\bm{\theta}_u,\bm{\rho}_u;\mathcal{X}_i)=\mathbb{E}_{q_{\bm{\phi}_u}(\bm{u}_i|\mathcal{X}_i)}[\log p_{\bm{\theta}_{u}}(\bm{R}_{i\cdot}|\bm{u}_i)] \nonumber  \\
&-\mathrm{KL}(q_{\bm{\phi}_u}(\bm{u}_i|\mathcal{X}_i)|| p_{\bm{\rho}_u}(\bm{u}_i|\bm{f}_i)),
\label{Eq:u}
\end{align}
\begin{align}
\mathcal{L}_{j}(q_{\bm{\phi}_v})=&\mathcal{L}(\bm{\phi}_v, \bm{\theta}_v, \bm{\rho}_v;\mathcal{Y}_j)=\mathbb{E}_{q_{\bm{\phi}_v}(\bm{v}_j|\mathcal{Y}_j)}[\log p_{\bm{\theta}_{v}}(\bm{R}_{\cdot j}|\bm{v}_j)]  \nonumber  \\
&-\mathrm{KL}(q_{\bm{\phi}_v}(\bm{v}_j|\mathcal{Y}_j) || p_{\bm{\rho}_v}(\bm{v}_j|\bm{g}_j)),
\label{Eq:v}
\end{align}
Since we assume the posteriors are Gaussian distribution, the $\mathrm{KL}$ terms in Eq.(\ref{Eq:u}) and Eq.(\ref{Eq:v}) have  analytical forms. However, for the expectation terms, we can not  compute them analytically. To handle this problem, we use  Monte Carlo method \cite{rezende2014stochastic} to approximate  the expectations by drawing samples from the posterior distribution. By using the reparameterization trick \cite{rezende2014stochastic}, for user $i$:
\begin{align}
\vspace{-1mm}
\mathcal{L}&(\bm{\phi}_u,\bm{\theta}_u, \bm{\rho}_u;\mathcal{X}_i) \simeq \tilde{\mathcal{L}}(\bm{\phi}_u,\bm{\theta}_u, \bm{\rho}_u;\mathcal{X}_i)=\label{loss2}  \displaybreak[3]\\
&\frac{1}{K}\sum_{k=1}^{K} (\log p_{\bm{\theta}_u}(\bm{R}_{i\cdot}|\bm{u}_{i}^k))
-\mathrm{KL}(q_{\bm{\phi}_u}(\bm{u}_i|\mathcal{X}_i)|| p_{\bm{\rho}_u}(\bm{u}_i|\bm{f}_i)),\nonumber
\end{align}
where $K$ is the size of the samplings, $\bm{u}_i^k=\bm{\mu}_{\bm{u}_i}'+\bm{\delta}_{\bm{u}_i}'\odot \bm{\epsilon}_{\bm{u}_i}^k$ with $\odot$ being an element-wise multiplication and $\bm{\epsilon}_{\bm{u}_i}^k$ being samples drawn from standard multivariate normal distribution. The superscript $k$ denotes the $k$-th sample.  The ELBO for item network, $\mathcal{L}(\bm{\phi}_v,\bm{\theta}_v, \bm{\rho}_v;\mathcal{Y}_j)$, can be derived similarly, and  we omit it here due to space limitations.

\begin{table*}[!htb]
\center
\caption{Statistics of datasets.}
\small
\vspace{-1.0em}
\label{table:datasets}
\resizebox{0.90\textwidth}{!}{
\begin{tabular}{c c c c c c c}
\toprule[0.8pt]
Dataset  & Item side information & User side information & \#Items & \#Users & \#Ratings& Sparsity \\
\hline
ML-100K &movie descriptions & tags & 1,682& 943& 100,000& 94.7\% \\
ML-1M & movie descriptions  & user demographics &  1,000,209 &  6,040 & 3,900 & 99.6\% \\
Lastfm-2K & tags  & social information  &  17,632 &  1,892 & 92,834 & 97.3\% \\
\bottomrule[0.8pt]
\end{tabular}}
\vspace{-0.5em}
\end{table*}

Since the expectation term in (Eq.\ref{Eq:fullloss}) is also parameterized by neural network, we can not solve it analytically. However, we notice we have sampled $\bm{u}_i^k$ and $\bm{u}_j^k$
when we solve $\mathcal{L}(\bm{\phi}_u,\bm{\theta}_u, \bm{\rho}_u;\mathcal{X}_i)$ and $\mathcal{L}(\bm{\phi}_v,\bm{\theta}_v, \bm{\rho}_v;\mathcal{Y}_j)$. The expectation term can also
be estimated by these samplings:
\begin{align}
\mathbb{E}&_{ q_{\bm{\phi}_u}(\bm{u}_i|\mathcal{X}_i)q_{\bm{\phi}_v}(\bm{v}_j|\mathcal{Y}_j)}
[\log p_\gamma(R_{ij}|\bm{u}_i,\bm{v}_j)] \simeq \nonumber \\
&\tilde{\mathcal{L}}_{ij}(\bm{\gamma};\bm{x}_i,\bm{y}_j) = \frac{1}{K}\sum_{k=1}^{K}\log p_{\bm{\gamma}}(R_{ij}|\bm{u}_{i}^k,\bm{v}_{j}^k),
\label{Eq:expection}
\end{align}
where $\bm{v}_j^k=\bm{\mu}_{\bm{v}_j}'+\bm{\delta}_{\bm{v}_j}'\odot \bm{\epsilon}_{\bm{v}_j}^k$.
\subsection{Optimization}
Since minimizing the objection function is equivalent to maximizing the conditional variational evidence lower bound (ELBO). Based on $\mathcal{L}(\bm{\phi}_u,\bm{\theta}_u, \bm{\rho}_u;\mathcal{X}_i)$ (i.e., Eq.\ref{loss2}), $\mathcal{L}(\bm{\phi}_v,\bm{\theta}_v, \bm{\rho}_v;\mathcal{Y}_j)$, Eq.\ref{Eq:expection} and Eq.\ref{Eq:cross-entropy}, the objective function can be represented as:
\begin{align}
\mathcal{L}=&-\sum_{i=1}^{M}\sum_{j=1}^{N}\sum_{k=1}^{K}\frac{1}{K} \log p_{\bm{\theta}_u}(\bm{R}_{i\cdot}|\bm{u}_{i}^k) + \log p_{\bm{\theta}_v}(\bm{R}_{\cdot j}|\bm{v}_{j}^k) \displaybreak[3] \nonumber\\
&+p_{\bm{\gamma}}(R_{ij}|\bm{u}_{i}^k,\bm{v}_{j}^k) -\mathrm{KL}(q_{\bm{\phi}_u}(\bm{u}_i|\mathcal{X}_i)|| p_{\bm{\rho}_u}(\bm{u}_i|\bm{f}_i)) \nonumber \\
&-\mathrm{KL}(q_{\bm{\phi}_v}(\bm{v}_j|\mathcal{Y}_j)|| p_{\bm{\rho}_v}(\bm{v}_j|\bm{g}_j)), \displaybreak[3]
\end{align}

\noindent Then, we can construct an estimator of the ELBO of the full dataset, based on minibatch:
\begin{align}
\mathcal{L}\simeq \tilde{\mathcal{L}}^{M}&=\frac{1}{E}\sum_{(i,j)}^{E}(\tilde{\mathcal{L}}(\bm{\phi}_u,\bm{\theta}_u, \bm{\rho}_u;\mathcal{X}_i)+ \\
&\tilde{\mathcal{L}}(\bm{\phi}_v,\bm{\theta}_v, \bm{\rho}_v;\mathcal{Y}_j)+\tilde{\mathcal{L}}_{ij}(\bm{\gamma};\bm{x}_i,\bm{y}_j)), \nonumber
\end{align}
where $E$ is the number of ($i$, $j$) positive  pairs ($R_{ij}=1$) sampled  from $\bm{R}$. Like mentioned in \cite{He:2017:NCF:3038912.3052569}, for negative feedback ($R_{ij}=0$), we can uniformly sample them from unoberved interactions in each iteration and control the negative sampling ratio ($neg\_ratio$). $E=E_p+E_n$, where $E_n=neg\_ratio\cdot E_p$ , $E_p$ and $E_{n}$ are the number of positive and negative sampling pairs, respectively.
Like mentioned in VAE \cite{Kingma2014Auto}, the number of samples $K$ per training pair  can be set to 1 as long as the minibatch size $E$ is large enough, e.g., $E$ = 128. Then we can update these parameters by using the gradient $\nabla_{\bm{\theta}_u, \bm{\phi}_u, \bm{\rho}_u, \bm{\theta}_v, \bm{\phi}_v, \bm{\rho}_v, \bm{\gamma}} \tilde{\mathcal{L}}^M$.
\subsection{Prediction}
After training, we can get the posterior distributions of  $\bm{u}_i$ and $\bm{v}_j$ through the user and item inference networks, respectively. So the prediction distribution $p(\hat{R}_{ij})$ can be made by:
\begin{align}
p(\hat{R}_{ij}|\bm{R})=\int p(\hat{R}_{ij}|\bm{u}_i,\bm{v}_j)q(\bm{u}_i|\bm{R})q(\bm{v}_j|\bm{R})d\bm{u}_i d\bm{v}_j. \label{Eq:cold1}
\end{align}
For a cold user, he/she has no previous feedback information. The posterior distribution $p(\bm{u}_i|\bm{R})$ is equal to the prior distribution $p(\bm{u}_i|\bm{f}_i)$. For cold user $i$, Eq.~\ref{Eq:cold1}  can be rewritten by:
\begin{align}
p(\hat{R}_{ij}|\bm{R})=\int p(\hat{R}_{ij}|\bm{u}_i,\bm{v}_j)p(\bm{u}_i|\bm{f}_i)q(\bm{v}_j|\bm{R})d\bm{u}_i d\bm{v}_j. \label{Eq:cold2}
\end{align}
For new item, the Eq.~\ref{Eq:cold1} can be rewritten similarly, and thus we omit here.  The integrals in Eq.~\ref{Eq:cold1} and Eq.~\ref{Eq:cold2} can't be solved analytically. We use the Monte Carlo approximation to the predict the $p(\hat{R}_{ij}|\bm{R})$
\begin{equation}
p(\hat{R}_{ij}|\bm{R})\approx \frac{1}{S}\sum_{s=1}^{S}p_{\bm{\gamma}}(R_{ij}|\bm{u}_{i}^s,\bm{v}_{j}^s),
\end{equation}
where $S$ is the number of samplings, $\bm{u}_i^k$ and $\bm{v}_j^k$ are  samplings from posterior distributions of  $\bm{u}_i$ and $\bm{v}_j$ (prior distribution in cold start scenario), respectively. Finally, we can use the expectation of $p(\hat{R}_{ij}|\bm{R})$ as the predictive value for user $i$ and item $j$.

\section{Experimental Setup}
\subsection{Research Questions}
We seek to answer these research questions that guide the remainder of the paper: 
{\bf (RQ1)} Does our proposed \VDMF {} outperform the state-of-the-art collaborative filtering methods for implicit feedback on real world sparse datasets? {\bf{(RQ2)}} Can our proposed model  effectively handle cold start problem? {\bf (RQ3)}  How do the key hyper-parameters ($neg\_ratio$ and embedding size $D$) of \VDMF{} affect recommendation performance? {\bf{(RQ4)}} Do the \emph{side ~information-specific} priors help to improve recommendation performance?
\subsection{Dataset} 
\textbf{MovieLens-100K} (ML-100k)\footnote{\textbf{\url{https://grouplens.org/datasets/movielens/100K/}}.}
. This dataset is a user-movie dataset. Each rating value is in range of 1-5. Since we consider implicit feedback, following ~\cite{dong2017hybrid,Li2015Deep}, we convert the  ratings $\geq 4$ to 1 and those $<4$ to 0. Same to \cite{Li2015Deep}, we regard users' tags as side information of the users. We use movie descriptions as side information of items.\\
\textbf{MovieLens-1M} (ML-1M)\footnote{
\textbf{\url{https://grouplens.org/datasets/movielens/1M/}}.}
. For this dataset, we convert ratings to implicit feedback as the same as we do for ML-1M. We take users' demographics (Gender, Age, Occupation and Zip code) as user side information and descriptions of movies as item side information.
\noindent \\
\textbf{Lastfm} (lastfm-2k)\footnote{
\textbf{\url{http://www.lastfm.com}}.} For this dataset, we set the user-item feedback $R_{ij}$ to be 1 if the user $i$ has listened to the item $j$; otherwise, it is 0.  We use items' tag information and users' social information as side information of items and users, respectively. Table~\ref{table:datasets} shows the statistics of the datasets and side information we used in the experiment. The three datasets have different sparsity ratios, which are for providing verification of model performance with different sparsities.
\subsection{Baselines and Experimental Settings} We make comparisons between our \VDMF{} and the following state-of-the-art baseline algorithms: \\
(1) \textbf{PMF}: This model \cite{mnih2008probabilistic} is a traditional latent factor model for CF. (2) \textbf{BPR}: This model~\cite{Rendle2009BPR} optimizes a pair-wise ranking loss for CF. (3) \textbf{DCF}: This model~\cite{Li2015Deep} uses marginalized denoising stacked auto-encoders to incorporate both users' and items' side information into MF. (4) \textbf{CVAE}: This model \cite{Li2017Collaborative} incorporates items side information into MF through VAE. (5) \textbf{NeuMF}\footnote{\textbf{\url{https://github.com/hexiangnan/neural_collaborative_filtering}}.}: This model \cite{He:2017:NCF:3038912.3052569} uses neural network to model interaction between users and items features. We use the source code provided by the author. (6) \textbf{VAE-CF}:  This model \cite{Liang:2018:VAC:3178876.3186150} is a state-of-the-art method that directly apply VAE to the task of CF. We use multinomial likelihood function in our experiment and set $\beta=0.1$. (7) \textbf{aSDAE}: This model \cite{dong2017hybrid} utilizes a hybrid stack denoising autoencoder to incorporate users' and items' side information into MF. (8) \textbf{NFM}: This model \cite{He2017Neural} is the state-of-the-art hybrid recommendation method, which uses Bi-Interaction layer to incorporate both feedback information and content information. We use pair-wise loss with negative sampling to train it. Similarly to \cite{He:2017:NCF:3038912.3052569,ijcai2017-447}, we adopt the leave-one-out evaluation method. For each user, we utilize the latest feedback for testing and the remaining data for training. Following \cite{He:2017:NCF:3038912.3052569,Elkahky:2015:MDL:2736277.2741667}, we also randomly choose 99 items which are not interacted by the user as negative samples for testing and
rank the test items among the 100 items for evaluation. For ItemKNN, We use the settings provided by \cite{Hu2009Collaborative}. For other baselines, the optimal parameters are set according to the literatures. For our \VDMF{}, we set $neg\_ratio=5$, $S=128$ and $D$=128. The minibatch is set to 128. The two generative networks both are two latent layers with Relu activation. The last layer of generative network is sigmod activation. The two prior networks are one latent layer. 
\begin{table*}
\center
\caption{Recommendation performance comparison between our \VDMF{} and baselines.}
\label{Table:performance}
\vspace{-1.0em}
\small
\begin{tabular}{cccccccccccc}
\toprule[1.0pt]
   Datasets           &     Metrics  &  PMF &   BPR &   DCF &  CVAE  &  NeuMF &VAE-CF& aSDAE & NFM &NVHCF  \\
\hline
\hline
 \multirow{2}{*}{ML-100K}
                           & HR@5  & 0.4634 & 0.4794 &  0.4708  & 0.4721 & 0.4942 & 0.5032& 0.4821 &  0.5057 & \textbf{0.5107} \\
                           & NDCG@5   & 0.3021 & 0.3188 & 0.3271  & 0.3185 & 0.3357 & 0.3401& 0.3298& 0.3398& \textbf{0.3508}\\

\hline
\hline
 \multirow{2}{*}{ML-1M}    & HR@5   & 0.5111 & 0.5312 &  0.5393  & 0.5392
                           & 0.5485 &    0.5642 & 0.5517 & 0.5621& \textbf{0.5725}\\
                           & NDCG@5   & 0.3463 & 0.3646 & 0.3690  & 0.3771 & 0.3865 & 0.3925& 0.3758 & 0.3886 & \textbf{0.4014}\\

\hline
\hline
 \multirow{2}{*}{Lastfm}
                           & HR@5   & 0.1214 & 0.1679 &  0.1627  & 0.1623
                           & 0.1701 &    0.1741 & 0.1587& 0.1677 &  \textbf{0.1786}\\
                           & NDCG@5   & 0.0617 & 0.0821 & 0.0889 & 0.0914  & 0.1078 & 0.1067 & 0.0987& 0.1082 & \textbf{0.1174}\\
\bottomrule[1.0pt]
\end{tabular}
\end{table*}
\subsection{Evaluation Metrics}
For evaluating our model's performance on implicit feedback, we use two common evaluation metrics for top-$k$ recommendation: HR$@k$ (Hit Radio at $k$)~\cite{He:2017:NCF:3038912.3052569} and NDCG$@k$ (Normalized Discounted Cumulative Gain at $k$)~\cite{croft2009search}.
\section{Results and Analysis}
\subsection{Overall Performance (RQ1)}
Table \ref{Table:performance} lists the top-$k$ recommendation performance  of  all methods on the three sparse datasets, ML-100K, ML-1M and Lastfm, in terms of HR$@5$ and NDCG$@5$.  The following findings can be observed from Table \ref{Table:performance}: (1) Most neural network-based algorithms, i.e., CVAE, NeuMF, VAE-CF and  \VDMF{}, outperform linear traditional baseline algorithms, e.g., PMF, which demonstrates that deep neural network does help to obtain more subtle and better latent representations of users and items. (2) Our \VDMF{} almost achieves all the best performance in terms of the three datasets and the two metrices, which confirms the effectiveness of our \VDMF{} to the CF task. (3) We also observe, the VAE-based method, i.e., VAE-CF and our \VDMF{} achieve promising  performance, which demonstrates that the Bayesian nature and non-linearity of neural network can help infer better latent preferences of users and items. (4) Although both based on VAE, our \VDMF{} outperforms  VAE-CF and CVAE in terms of all datasets and all metrices, which shows the advantage of our conditional VAE framework. (5) Our \VDMF{} outperforms  state-of-the-art hybrid methods including DCF, CVAE, aSDAE and NFM, which demonstrates the effectiveness of  our way of incorporating side information.
\par
\begin{figure}
\centering
  \subfigure{
    \label{subfig:neg_ratio}
    \includegraphics[width=0.22\textwidth]{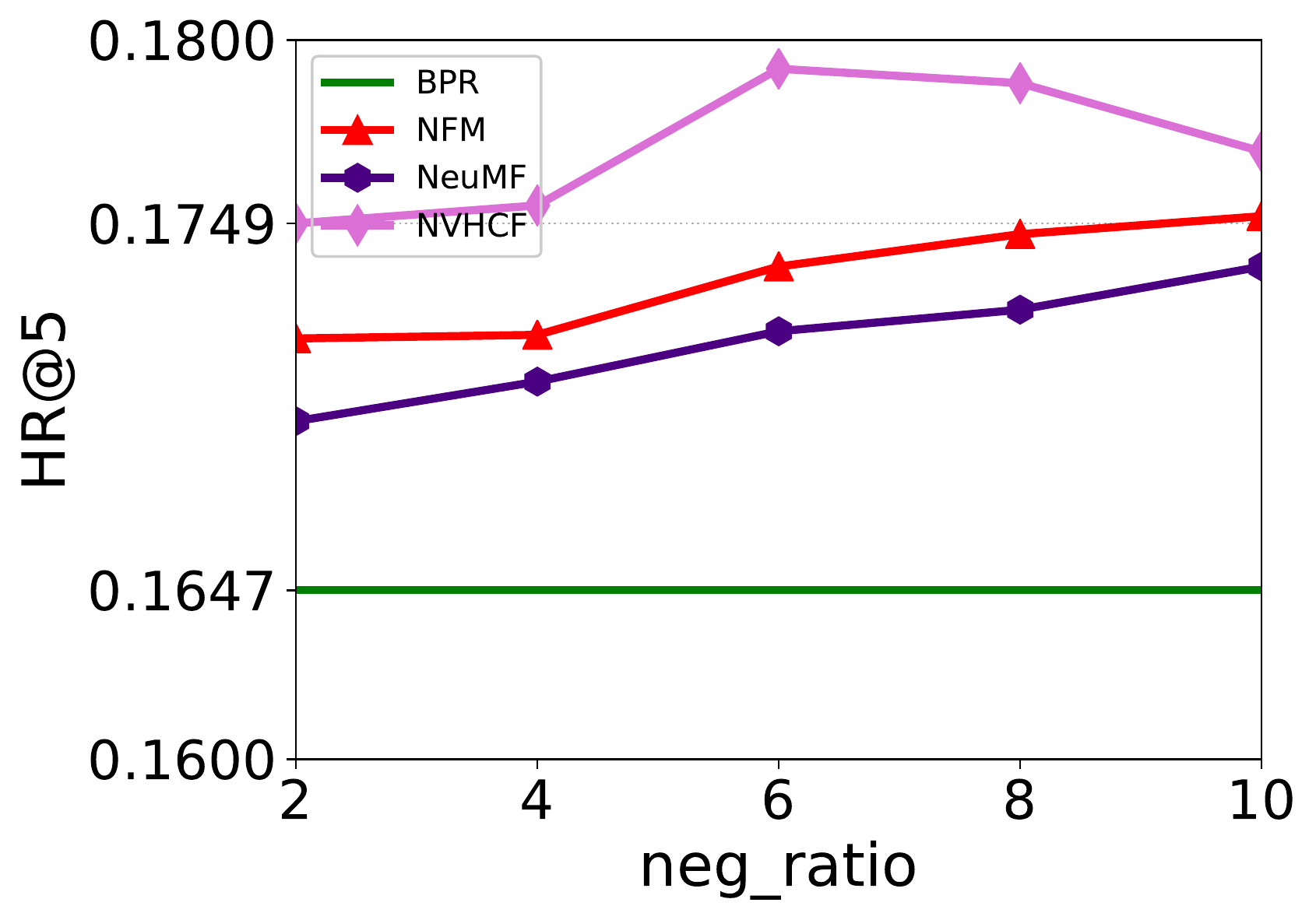}}
  \subfigure{
    \label{subfig:neg_ratio}
    \includegraphics[width=0.22\textwidth]{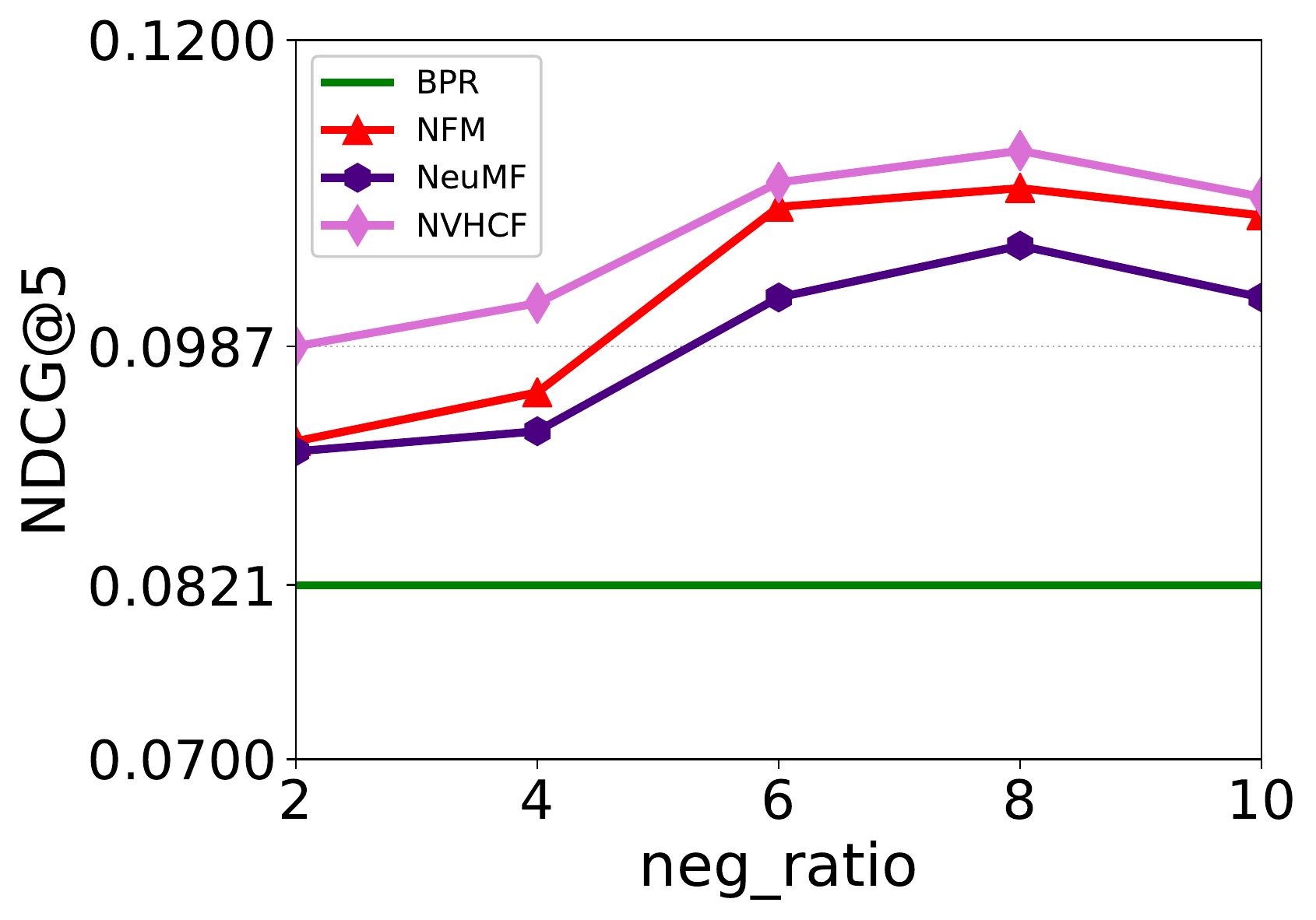}}
      \vspace{-1.5em}
\caption{Performance on HR$@5$ and NDCG$@5$ metrics with different negative sampling ratio on Lastfm dataset.}\label{fig:negative-Lastfm}
\end{figure}
\begin{figure}
\centering
  \subfigure{
    \label{subfig:neg_ratio}
    \includegraphics[width=0.22\textwidth]{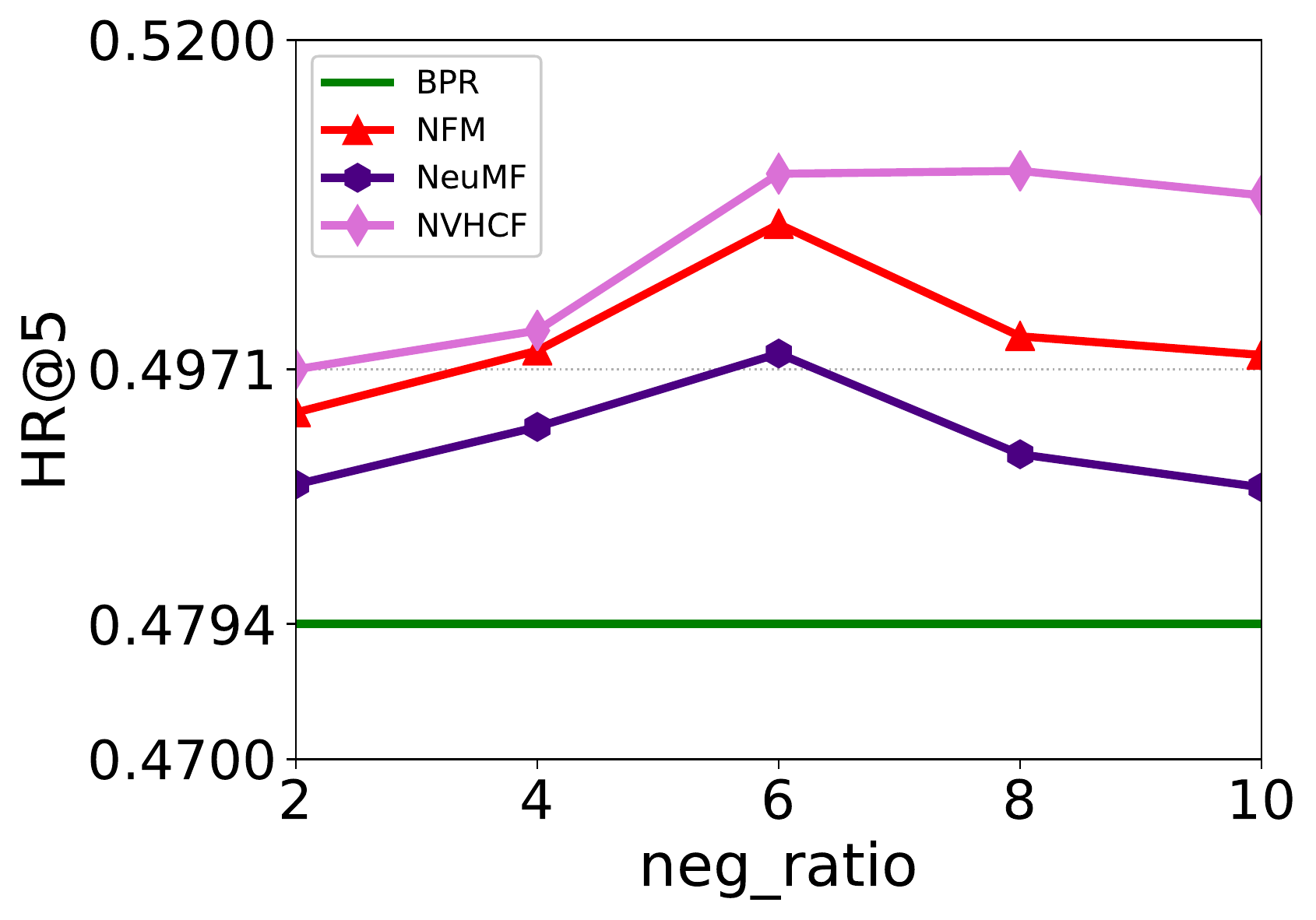}}
  \subfigure{
    \label{subfig:neg_ratio}
    \includegraphics[width=0.22\textwidth]{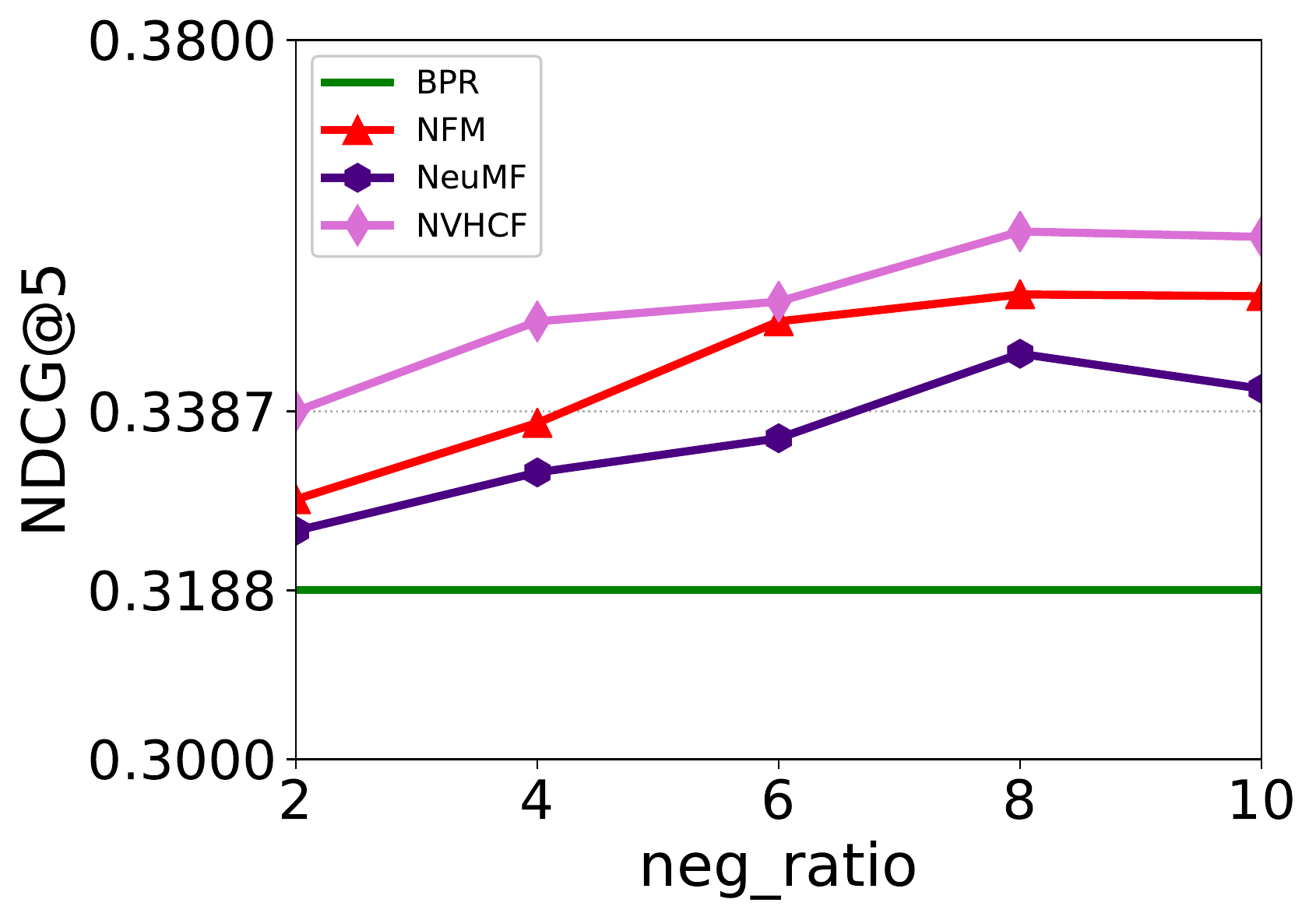}}
      \vspace{-1.5em}
\caption{Performance on HR$@5$ and NDCG$@5$ metrics with different negative sampling ratio on ML-100K dataset.}\label{fig:negative-ML100K}
\end{figure}
\subsection{Cold Start Performance Comparison (RQ2)}
To evaluate our model on different cold start scenarios, similar to \cite{shi2018attention}, we form evaluation sets in different cold ratios. We first split the dataset into training ($80\%$), validation ($10\%$) and test sets ($10\%$). For $30\%$ cold users, we random choose $30\%$ samples  in the validation and test sets and give each sample a specific user id only for the sample. We evaluate our model in $30\%$, item cold (cold-i), $30\%$ user cold (cold-u)  scenarios on three datasets in term of NDCG$@$5. Since some baselines (BPR, NeuMF and VAE-CF) only use feedback information and don't work properly on cold scenario, we don't compare \VDMF{} with them. Tabel.~\ref{Table:cold} shows the performance of \VDMF{} and other hybrid methods in different cold start scenarios. Note that  CVAE cannot handle cold user problem, thus we don't report experiments of it in cold user scenario. As it can be seen, \VDMF{} significantly  outperforms recent hybrid methods, CVAE, aSDAE, NFM in the scenarios of both cold items and cold users, which illustrates that latent prior representations generated by \VDMF{} in cold scenarios  work better than the state-of-the-art. The finding that \VDMF{} and NFM outperform CVAE, aSDAE and DCF indicates that using neural network to model interactions between users and items works better than those of simply using dot product.
\begin{table}
\center
\caption{Recommendation performance comparison in different cold start scenarios on three datasets in terms of NDCG$@$5.}
\label{Table:cold}
\vspace{-1em}
\begin{tabular}{ccccccccc}
\toprule[1.0pt]
  & \multicolumn{2}{c}{ML-100K} & & \multicolumn{2}{c}{ML-1M} & \space & \multicolumn{2}{c}{Lastfm} \\
  \cline{2-3}  \cline{5-6} \cline{8-9}
  & cold-i & cold-u && cold-i & cold-u & & cold-i & cold-u \\
  \hline
  \hline
 DCF & .1441 & .1659 & & .1881& .1922& & .0531& .0362\\
 CVAE &.1385&-&&.1571&-&&.0414&-\\
 aSDAE &.1762&.1726&&.1865&.2124&&.0734&.0525\\
 NFM &.1778&.1823&&.1953&.2274&&.0921&.0603\\
 NVHCF &\textbf{.1927}&\textbf{.2033}&&\textbf{.2151}&\textbf{.2435}&&\textbf{.1124}&\textbf{.0741}\\
\bottomrule[1.0pt]
\end{tabular}
\end{table}
\begin{figure}
\centering
  \subfigure{
    \label{subfig:dimensions-HR}
    \includegraphics[width=0.22\textwidth]{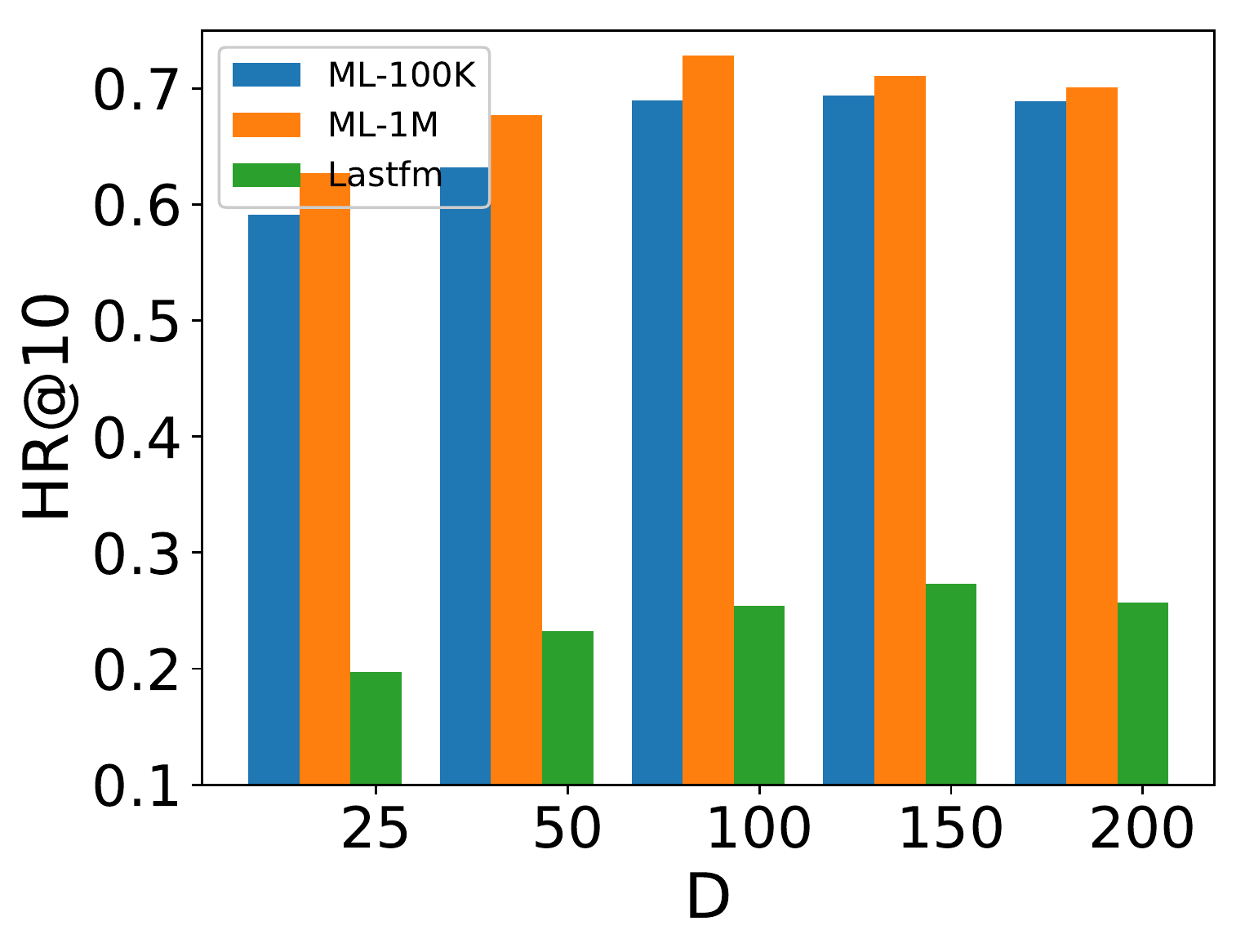}}
  \subfigure{
    \label{subfig:dimensions-NDCG}
    \includegraphics[width=0.22\textwidth]{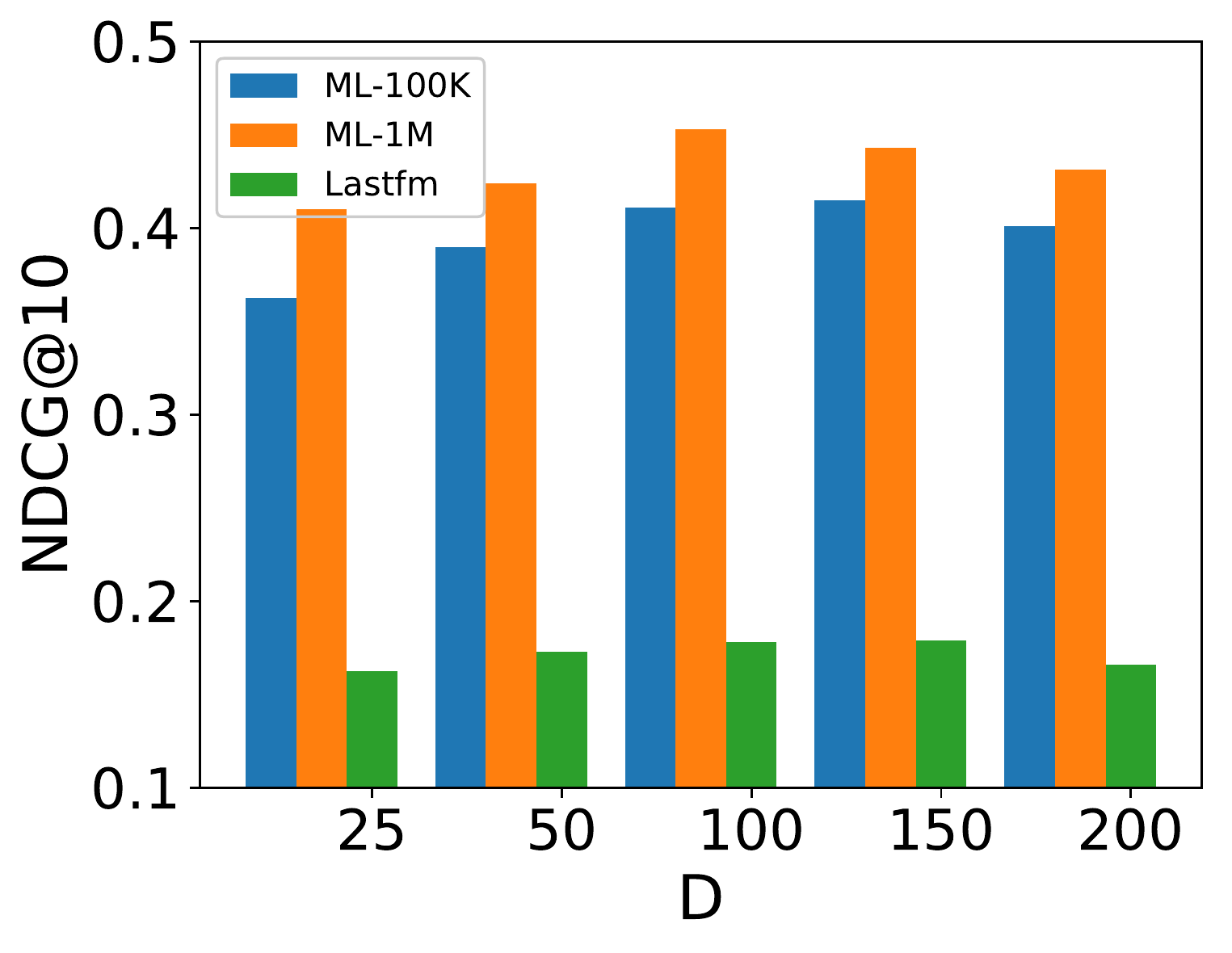}}
    \vspace{-1.5em}
\caption{Performance on HR$@10$ and NDCG$@10$ metrics with different embedding size on the three datasets.}\label{fig:dimensions}
\end{figure}
\subsection{Sensitivity Analysis (RQ3)} Next, we turn to answer research question \textbf{RQ3}. We study the effect of the hyper-parameters on recommendation performance.
To evaluate the effects of the dimension of latent space, we compare the performance for different dimensions fixing $neg\_ratio=5$  parameters on the three datasets in term of NDCG$@10$ and HR$@10$, where the size of embeddings' dimension, $D$, is set to be 25,50,100,150 and 200, respectively. We can observe that larger dimension leads to better performance. Specifically, the optimal embedding size of \VDMF{} for Lastfm is 150, and for ML-100K and ML-1M it is 100. According to Figure \ref{fig:dimensions}, our \VDMF {} outperforms other baselines on different embedding size. This, once again, demonstrates the effectiveness of out \VDMF{} for top-$k$ recommendation.\par
To understand the influence of the negative sampling ratio ($neg\_r$
$atio$$\in[2,4,6,8,10]$) on \VDMF{} and other baselines which involve negative sampling (e.g., NeuMF, NFM and BPR), we compare the performance for different $neg\_ratio$ on ML-100K and Lastfm in terms of HR$@5$ and NDCG$@5$. It should be noted that the $neg\_ratio$ is fixed to 1 for BPR due to its pairwise objection \cite{He:2017:NCF:3038912.3052569}. Figures \ref{fig:negative-ML100K} and \ref{fig:negative-Lastfm} show the performance \emph{w.r.t} different negative sampling ratios. The following findings can be observed from Figures \ref{fig:negative-ML100K} and \ref{fig:negative-Lastfm}: (1) In general, sampling more negative samples will lead to better performance. (2) \VDMF{} beats other baselines on the two datasets. (3) For the two datasets, the optimal ratio for our \VDMF{} is between $6$ and $8$, which indicates we can tune $neg\_ratio$ to achieve best performance.\\
\subsection{Contribution of \textit{Side~Information-Specific} Priors (RQ4)}
Finally, we turn to answer \textbf{RQ4} for understanding  the effect of \emph{side~ information-specific}. We consider three variants of our \VDMF{}: NVH-n (NVHCF-none) , NVH-u (NVHCF-user) and NVH-i (NVHCF-item). For example, the  NVHCF-item represents we only keep the item prior network and remove the user prior network  (the KL divergence $\mathrm{KL}(q_{\bm{\phi}_u}(\bm{u}_i|\mathcal{X}_i)||p_{\bm{\rho}_u}(\bm{u}_i|\bm{f}_i))$ in Eq.\ref{loss2} degenerate to $\mathrm{KL}(q_{\bm{\phi}_u}(\bm{u}_i|\mathcal{X}_i)||\mathcal{N}(0,\bm{I}_D))$, which means the priors for all users are the same standard Normal distribution. Similarly, the NVHCF-n represents the model where we remove both users and items prior networks. Table \ref{Table:variants} shows recommendation performance between different variants of \VDMF{}. We can find \VDMF{} outperforms other variants, which demonstrates considering both users' and items' \emph{side~information-speific} can get better performance than considering only one of them. We can observe NVHCF-i betters NVHCF-u on Movie datasets (ML-100K,ML-1M), which demonstrates that incorporating items side information into prior is more effective than incorporating users side information. This may be due to the fact that the movies' side information (features) can better model its latent represenation than users' side information. In contrast, for Lastfm dataset, the users' side information (social information) is more helpful than items' side information to improve recommendation performance.
\begin{table}
\center
\caption{Recommendation performance comparison between different variants of  \VDMF{} in terms of HR$@$10 and NDCG$@$10.}
\vspace{-1.0em}
\label{Table:variants}
\begin{tabular}{cccccc}
\toprule[1.0pt]
   Datasets           &     Metrics  &   NVH-n  &  NVH-u& NVH-i & NVH \\
\hline
\hline
 \multirow{2}{*}{ML-100K}
                           & HR@10   & .6531 & .6611 &  .6734  & \textbf{.6896} \\
                           & NDCG@10   & .3986 & .4017 & .4052  & \textbf{.4108} \\
\hline
\hline

\multirow{2}{*}{ML-1M}     & HR@10   & .6952 &  .7178  &  .7231  & \textbf{.7286} \\
                           & NDCG@10   & .4128 & .4412 & .4431  & \textbf{.4528} \\
\hline
\hline
\multirow{2}{*}{Lastfm}    & HR@10   & .2438 &  .2512  &  .2481  & \textbf{.2543} \\
                           & NDCG@10   & .1644 & .1725 & .1689  & \textbf{.1782} \\

\bottomrule[1.0pt]
\end{tabular}
\vspace{-0.5em}
\end{table}


\section{Conclusions}
In this paper, we studied the problem of inferring effective latent factors of users and items for CF. We have proposed a new algorithm, Neural Variational Hybrid Collaborative Filtering, \VDMF{}, that is the first unified deep generative framework for hybrid collaborative filtering. Our \VDMF{} models both users' and items' generative processes, which enables it to make recommendation when a new user or a new item comes. Our \VDMF{} incorporates side information of users and items through  \emph{side ~information-specific} priors, which enables our model to alleviate matrix sparsity and better model users' preference and items' features. For inference, we proposed a conditional stochastic gradient variational bayesian algorithm. The Bayesian nature and non-linearity of the neural network enable our \VDMF{} to learn better latent factors of users and items. Our \VDMF{} is a unified deep generative model which make it be able to  handle the cold start problem via a full Bayesian probabilistic view.
Experimental results show that our \VDMF{} yields better recommendation performance and effectively handles cold start problem.
As to future work, we plan to apply \VDMF{} to tackle other recommendation tasks such as recommending a knowledgeable user to a question in question-answering community.


\bibliographystyle{ACM-Reference-Format}
\bibliography{references}

\end{document}